\documentclass[submitting]{nst}

\usepackage{subfigure,dcolumn}
\usepackage[T2A,T1]{fontenc}
\usepackage[russian,english]{babel}

\usepackage{listings}
\begin{document}

\title{Huizhou Hadron Spectrometer - a Proposed High-rate Experimental Setup at the High Intensity Heavy-ion Accelerator Facility}

\thanks{Supported by the CAS Project for Young Scientists in Basic Research (No. YSBR-088), the National Science Foundation of China (Nos. 12347103, 
12435007, 12361141819, 
12305145) 
and Chongqing Natural Science Foundation (No. CSTB2025NSCQ-GPX0745). 
}

\author{Xurong Chen}
\affiliation{State Key Laboratory of Heavy Ion Science and Technology, Institute of Modern Physics, Chinese Academy of Sciences, Lanzhou 730000, China}
\affiliation{School of Nuclear Science and Technology, University of Chinese Academy of Sciences, Beijing 100049, China}
\affiliation{Southern Center for Nuclear-Science Theory (SCNT), Institute of Modern Physics, Chinese Academy of Sciences, Huizhou 516000, China}

\author{Yunyun Fan}
\affiliation{Institute of High Energy Physics, Chinese Academy of Sciences, Beijing 100049, China}
\affiliation{School of Physical Sciences, University of Chinese Academy of Sciences, Beijing 100049, China}

\author{Shuangshi Fang}
\affiliation{Institute of High Energy Physics, Chinese Academy of Sciences, Beijing 100049, China}
\affiliation{School of Physical Sciences, University of Chinese Academy of Sciences, Beijing 100049, China}
\affiliation{Center for High Energy Physics, Henan Academy of Sciences, Zhengzhou 450046, China}

\author{Zhaoqing Feng}
\affiliation{School of Physics and Optoelectronics, South China University of Technology, Guangzhou 510640, China}
\affiliation{State Key Laboratory of Heavy Ion Science and Technology, Institute of Modern Physics, Chinese Academy of Sciences, Lanzhou 730000, China}

\author{Fengkun Guo}
\affiliation{Institute of Theoretical Physics, Chinese Academy of Sciences, Beijing 100190, China}
\affiliation{School of Physical Sciences, University of Chinese Academy of Sciences, Beijing 100049, China}
\affiliation{Southern Center for Nuclear-Science Theory (SCNT), Institute of Modern Physics, Chinese Academy of Sciences, Huizhou 516000, China}

\author{Weijia Han}
\affiliation{State Key Laboratory of Heavy Ion Science and Technology, Institute of Modern Physics, Chinese Academy of Sciences, Lanzhou 730000, China}

\author{Jun He}
\affiliation{Department of Physics, Nanjing Normal University, Nanjing 210023, China}

\author{Qinghua He}
\affiliation{Department of Nuclear Science and Engineering, Nanjing University of Aeronautics and Astronautics, Nanjing 210016, China}

\author{Xionghong He}
\affiliation{State Key Laboratory of Heavy Ion Science and Technology, Institute of Modern Physics, Chinese Academy of Sciences, Lanzhou 730000, China}
\affiliation{School of Nuclear Science and Technology, University of Chinese Academy of Sciences, Beijing 100049, China}

\author{Hongxia Huang}
\affiliation{Department of Physics, Nanjing Normal University, Nanjing 210023, China}

\author{Xiaolin Kang}
\affiliation{School of Mathematics and Physics, China University of Geosciences, Wuhan 430074, China}

\author{Hongli Ma}
\affiliation{Department of Physics, Chongqing Key Laboratory for Strongly Coupled Physics, Chongqing University, Chongqing 401331, China}

\author{Weihu Ma}
\affiliation{Institute of Modern Physics, Fudan University, Shanghai 200433, China}

\author{Yongliang Ma}
\affiliation{School of Frontier Sciences, Nanjing University, Suzhou, 215163, China}
    
\author{Norihito Muramatsu}    
\affiliation{State Key Laboratory of Heavy Ion Science and Technology, Institute of Modern Physics, Chinese Academy of Sciences, Lanzhou 730000, China}
    
\author{Zaiba Mushtaq}
\affiliation{State Key Laboratory of Heavy Ion Science and Technology, Institute of Modern Physics, Chinese Academy of Sciences, Lanzhou 730000, China}
\affiliation{School of Nuclear Science and Technology, University of Chinese Academy of Sciences, Beijing 100049, China}

\author{Xiaoyang Niu}
\affiliation{State Key Laboratory of Heavy Ion Science and Technology, Institute of Modern Physics, Chinese Academy of Sciences, Lanzhou 730000, China}
\affiliation{School of Nuclear Science and Technology, University of Chinese Academy of Sciences, Beijing 100049, China}

\author{Jialun Ping}
\affiliation{Department of Physics, Nanjing Normal University, Nanjing 210023, China}

\author{Hao Qiu}
\affiliation{State Key Laboratory of Heavy Ion Science and Technology, Institute of Modern Physics, Chinese Academy of Sciences, Lanzhou 730000, China}
\affiliation{School of Nuclear Science and Technology, University of Chinese Academy of Sciences, Beijing 100049, China}

\author{Jun Shi}
\affiliation{State Key Laboratory of Nuclear Physics and Technology, Institute of Quantum Matter, South China Normal University, Guangzhou 510006, China}
\affiliation{Guangdong Basic Research Center of Excellence for Structure and Fundamental Interactions of Matter, Guangdong Provincial Key Laboratory of Nuclear Science, Guangzhou 510006, China}

\author{Ye Tian}
\affiliation{State Key Laboratory of Heavy Ion Science and Technology, Institute of Modern Physics, Chinese Academy of Sciences, Lanzhou 730000, China}
\affiliation{School of Nuclear Science and Technology, University of Chinese Academy of Sciences, Beijing 100049, China}

\author{Qian Wang}
\affiliation{State Key Laboratory of Nuclear Physics and Technology, Institute of Quantum Matter, South China Normal University, Guangzhou 510006, China}
\affiliation{Guangdong Basic Research Center of Excellence for Structure and Fundamental Interactions of Matter, Guangdong Provincial Key Laboratory of Nuclear Science, Guangzhou 510006, China}

\author{Rong Wang}
\affiliation{State Key Laboratory of Heavy Ion Science and Technology, Institute of Modern Physics, Chinese Academy of Sciences, Lanzhou 730000, China}
\affiliation{School of Nuclear Science and Technology, University of Chinese Academy of Sciences, Beijing 100049, China}

\author{Shuaichun Wang}
\affiliation{Institute of Modern Physics, Fudan University, Shanghai 200433, China}

\author{Xiaoyun Wang}
\affiliation{Department of Physics, Lanzhou University of Technology,
Lanzhou 730050, China}

\author{Jiajun Wu}
\affiliation{School of Physical Sciences, University of Chinese Academy of Sciences, Beijing 100049, China}
\affiliation{Southern Center for Nuclear-Science Theory (SCNT), Institute of Modern Physics, Chinese Academy of Sciences, Huizhou 516000, China}

\author{Chuwen Xiao}
\affiliation{Department of Physics, Guangxi Normal University, Guilin 541004, China}

\author{Jujun Xie}
\affiliation{State Key Laboratory of Heavy Ion Science and Technology, Institute of Modern Physics, Chinese Academy of Sciences, Lanzhou 730000, China}
\affiliation{School of Nuclear Science and Technology, University of Chinese Academy of Sciences, Beijing 100049, China}
\affiliation{Southern Center for Nuclear-Science Theory (SCNT), Institute of Modern Physics, Chinese Academy of Sciences, Huizhou 516000, China}

\author{Haibo Yang}
\affiliation{State Key Laboratory of Heavy Ion Science and Technology, Institute of Modern Physics, Chinese Academy of Sciences, Lanzhou 730000, China}
\affiliation{School of Nuclear Science and Technology, University of Chinese Academy of Sciences, Beijing 100049, China}

\author{Xieyang Yu}
\affiliation{State Key Laboratory of Heavy Ion Science and Technology, Institute of Modern Physics, Chinese Academy of Sciences, Lanzhou 730000, China}
\affiliation{School of Nuclear Science and Technology, University of Chinese Academy of Sciences, Beijing 100049, China}

\author{Honglin Zhang}
\affiliation{State Key Laboratory of Heavy Ion Science and Technology, Institute of Modern Physics, Chinese Academy of Sciences, Lanzhou 730000, China}

\author{Shenghui Zhang}
\affiliation{Department of Physics, Chongqing Key Laboratory for Strongly Coupled Physics, Chongqing University, Chongqing 401331, China}

\author{Chengxin Zhao}
\affiliation{State Key Laboratory of Heavy Ion Science and Technology, Institute of Modern Physics, Chinese Academy of Sciences, Lanzhou 730000, China}

\author{Kuangta Chao}
\affiliation{School of Physics and Center for High Energy Physics, Peking University, Beijing 100871, China}

\author{Qiang Zhao}
\affiliation{Institute of High Energy Physics, Chinese Academy of Sciences, Beijing 100049, China}

\author{Bingsong Zou}
\affiliation{Department of Physics and Center for High Energy Physics, Tsinghua University, Beijing 100084, China}

\begin{abstract}

The High-Intensity Heavy-Ion Accelerator Facility (HIAF), currently under construction in Huizhou, Guangdong Province, China, is projected to be completed by 2025. 
This facility will be capable of producing proton and heavy-ion beams with energies reaching several GeV, thereby offering a versatile platform for advanced fundamental physics research.
Key scientific objectives include exploring physics beyond the Standard Model through the search for novel particles and interactions, testing fundamental symmetries, investigating exotic hadronic states such as di-baryons, pentaquark states and multi-strange hypernuclei, conducting precise measurements of hadron and hypernucleus properties, and probing the phase boundary and critical point of nuclear matter. 
To facilitate these investigations, we propose the development of a dedicated experimental apparatus at HIAF — the Huizhou Hadron Spectrometer (HHaS). 
This paper presents the conceptual design of HHaS, comprising a solenoid magnet, a five-dimensional silicon pixel tracker, a Low-Gain Avalanche Detector (LGAD) for time-of-flight measurements, and a Cherenkov-scintillation dual-readout electromagnetic calorimeter. 
The design anticipates an unprecedented event rate of 1–100 MHz, extensive particle acceptance, a track momentum resolution at 1$\%$ level, an electromagnetic energy resolution of $\sim$3$\%$ @ 1 GeV and multi-particle identification capabilities. 
Such capabilities position HHaS as a powerful instrument for advancing experimental studies in particle and nuclear physics. 
The successful realization of HHaS is expected to significantly bolster the development of medium- and high-energy physics research within China.
\end{abstract}

\keywords{High-Intensity Heavy-Ion Accelerator Facility, physics beyond the standard model, $\eta$ meson physics, light hadron physics.}

\maketitle

\section{Introduction}
\label{sec:introduction}

The High-Intensity Heavy-Ion Accelerator Facility (HIAF)~\cite{Yang:2013yeb, Zhou:2022pxl}, one of China's major national scientific and technological infrastructures, is currently under construction in Huizhou city, Guangdong province, with completion anticipated by 2025. 
Designed to deliver cutting-edge capabilities, HIAF will provide intense proton ($p$) and heavy-ion (e.g., uranium) beams with energies up to 9.3 GeV and 2.45 GeV/u, respectively, at its high energy terminal.
Following a planned upgrade, the facility will further enhance its performance, enabling the acceleration of uranium beams to energies as high as 9.1 GeV/u. 
Secondary $\pi$, $K$ meson beams and a polarized $p$ beam can also be provided with future upgrades.
This state-of-the-art facility, serving as a nuclear physics research center, offers unprecedented opportunities to explore the fundamental properties of matter and the dynamics of strong interactions~\cite{An_2025}.

The high-intensity $p$, $\pi$ and $K$ beams will produce vast quantities of light hadrons, composed by $u$, $d$, and $s$ quarks, enabling detailed studies of their decay patterns, mass spectra, and internal structures, shedding light on the mechanism of quark confinement, which is a key property of the strong interaction.
It will also provide opportunities to search for new particles and interactions beyond the Standard Model (SM)~\cite{Holdom:1985ag, Galison:1983pa, Fayet:1990wx, Burgess:2000yq, OConnell:2006rsp, Georgi:1986df, Aloni:2018vki, Landini:2019eck}, 
to test fundamental symmetries like CP symmetry~\cite{KLOE:2008arm, Gardner:2019nid, Shi:2024yfa, Sanchez-Puertas:2018tnp}, 
as well as to search for new exotic hadron states~\cite{PhysRevLett.96.042002, Wu:2023ywu, He:2017aps, Wang:2024qnk, Wang:2024xvq} and di-baryons~\cite{Clement:2016vnl, Dong:2017olm}.
On the other hand, heavy-ion beams at HIAF can be used to search for and locate the first-order phase boundary with the critical point which is predicted to exist between the hadron gas and quark-gluon plasma states at very high baryon density.
Hypernuclei, a strongly interacting many-body system composed of nucleons and hyperons, can also be produced and studied in detail with the heavy-ion beams, shedding light on the hyperon-nucleon and hyperon-hyperon interactions, which are crucial to understand the inner structure of neuron stars.

To explore these hadron-related physics subjects with unparalleled precision, we propose a new experimental setup, Huizhou Hadron Spectrometer (HHaS), at HIAF.
Envisioned as a light hadron factory, HHaS aims to revolutionize the field of particle and nuclear physics by achieving an unprecedented increase in the statistical precision — by $\sim$2 orders of magnitude — compared to existing experimental setups (e.g., KLOE-2~\cite{KLOE:1993sge}, BESIII~\cite{BESIII:2009fln}). 
The successful construction and operation of HHaS will significantly promote a broad range of sub-atomic physics research and advance our understanding of the nature.

The following sections of this paper describe the potential physics research directions, the performance requirements, the conceptual design and the expected performances of HHaS.

\section{Physics}
\label{sec:physics}

While the SM has been extremely successful in the past several decades in describing the most basic building blocks in our Universe and the interactions among them, there are several phenomena calling for new physics beyond SM~\cite{Essig:2013lka, Bertone:2004pz}. 
For example, the small but finite neutrino masses~\cite{ParticleDataGroup:2024cfk}, the existence of dark matter~\cite{Drees:2017xed, Banik:2023yfh}, and the matter-antimatter asymmetry in the Universe~\cite{ParticleDataGroup:2024cfk} all point to the need for new theories. 
Thus, the search for Beyond-Standard-Model (BSM) physics is one of the most exciting frontiers of particle physics~\cite{ParticleDataGroup:2024cfk}. 
The $\eta$ meson, a pseudo-scalar meson with all-zero discrete quantum numbers, i.e., $Q=I=J=S=B=L=0$, is an excellent probe to BSM physics \cite{Gan:2020aco}. 
All of its possible strong decays are forbidden in the lowest order by $P$, $CP$, charge, isospin and $G$-parity conservation.
Its EM decays are forbidden in the lowest order by $C$ invariance and angular momentum conservation.
As a result, BSM decay fractions of the $\eta$ meson are enhanced, offering potential avenues to search for dark portal particles like dark photons, dark Higgs, and axion-like particles~\cite{Gan:2020aco,Gao:2024rgl,Ding:2024iqr,Alonso-Alvarez:2023wni,
Holdom:1985ag,Galison:1983pa,Fayet:1990wx,Burgess:2000yq,OConnell:2006rsp,Georgi:1986df,Aloni:2018vki,Landini:2019eck}. 
Additionally, the detailed analysis of $\eta$ meson decays can help test fundamental symmetries, 
such as C, CP symmetries \cite{KLOE:2008arm,Gardner:2019nid,Shi:2024yfa,Sanchez-Puertas:2018tnp} and lepton flavor conservation \cite{White:1995jc}.
The decay of $\eta$ meson has been studied in various experiments in $e^+e^-$, $e^-p$, and $pp$ 
collisions~\cite{WASA-at-COSY:2018jdv,LHCb:2023iyw,LHCb:2016hxl,KLOE:2008arm,BESIII:2021fos,BESIII:2015fid,A2:2017gwp}. 
The REDTOP collaboration~\cite{REDTOP:2022slw} proposes an $\eta$ factory experiment using $pA$ or $\pi^{+}A$ collisions with a fixed nuclear target to conduct the measurements of the $\eta$ meson decays with a new level of precision.
With the high-intensity proton and $\pi^{+}$ beams available at HIAF, HHaS can also function as an $\eta$ factory.
The large $\eta$ production cross section in hadronic collisions, combined with the high luminosity easily obtained in fixed-target experiments, offers a great potential to produce $\eta$ samples with orders of magnitude higher statistics than the current data~\cite{Chen:2024wad}.

Due to the non-perturbative nature of quantum chromodynamics (QCD) at low energies, our understanding of the strong interaction is still limited.
Understanding how quarks are confined in hadrons remains one of the central challenges in modern physics. 
Hadron spectroscopy, the analysis of the excitation spectrum of the particles, has become a focal point for shedding light on the interaction and dynamics between the quarks, 
spurring major experimental efforts at facilities such as ELSA, CEBAF, ESRF, SPring-8, BEPCII, SuperKEKB, JPARC, GSI, COSY, SPS and LHC~\cite{PhysRevD.110.030001, THIEL2022103949, WANG:2025fmh, Mura2023PRC107}.
The proton beam and the secondary $\pi$ and $K$ meson beams at HIAF are able to provide a wealth of experimental data for studying light hadron spectroscopy and related dynamics. 
First, the high-luminosity proton beam can be used to study baryon excitations, hyperons and dibaryon states~\cite{Clement:2016vnl, Zou:2009wp, Dong:2017olm} by bombarding proton or nuclear targets.
Especially, as partner states of the $P_c$ pentaquark states, the $P_s$ states with hidden strangeness quark pair can also be searched for in the final state $p\phi$ in $pp$ collisions ~\cite{PhysRevLett.96.042002,Wu:2023ywu}.
Additionally, the hyperon polarization in unpolarized collision systems remains a long-standing question in hyperon production. 
The study of the $K\Lambda N$ final state offers a unique opportunity to investigate this issue and elucidate the inner structure of the $\Lambda$ hyperon.
Second, the secondary $\pi$ beam can provide more precise $\pi N$ scattering data, especially for final states such as $\pi N \to K\Lambda$, $K\Sigma$, and $\omega N$~\cite{Ronchen:2012eg, Kamano:2013iva, Wang:2022osj}. 
The uncertainties in the differential cross-sections of these two-body final states remain relatively large, and theoretical models have not yet reached a consensus~\cite{Wang:2022osj, Matsuyama:2006rp}.
Consequently, the extracted parameters for $N^*$ and $\Delta^*$ particles vary significantly. 
Therefore, more precise experimental data are urgently needed.
Exotic hadron states, like the pentaquark state mentioned above, may also be searched for in $\pi$p scatterings ~\cite{PhysRevLett.96.042002,Wang:2024qnk,Wang:2024xvq}.
Finally, the $K$ beam, which is expected to exceed the energy range and luminosity of the J-PARC~\cite{Aoki:2021cqa} and JLab~\cite{KLF:2020gai} experiments, will provide an ideal platform for studying hyperon excitations ~\cite{Kamano:2015hxa, Wu:2009tu}. 
Notably, the number of $\Xi$ particle excitations listed in the PDG is currently fewer than 10~\cite{ParticleDataGroup:2024cfk}, which is far fewer than the predictions of the quark model. 
At HIAF, more $\Xi$ excitations can be discovered through the $KN \to \bar{K} \Xi^*$ reaction~\cite{Guo:2025gba, Guo:2023gvx}.
For most of the above mentioned studies, polarization observables measured with polarized targets and / or beams provide valuable constraints on partial wave analysis in disentangling overlapping baryon resonances~\cite{Gao:2012zh,Shi:2014vha}. 

After the Quark-Gluon Plasma (QGP) has been created and studied in relativistic heavy-ion collisions for two decades~\cite{STAR:2005gfr, PHENIX:2004vcz, PHOBOS:2004zne, BRAHMS:2004adc}, a new frontier in high energy nuclear physics is searching for the predicted first-order phase transition and critical point between the QGP and hadronic phases in the high baryon density region of the QCD phase diagram~\cite{PhysRevD.101.054032}. 
Experimentally, this requires lower collision energies than the top RHIC and LHC energies~\cite{thestarcollaboration2025precisionmeasurementnetprotonnumber}.
A beam energy scan program has been carried out at RHIC, and new facilities like NICA~\cite{Golovatyuk:2016zps}, FAIR~\cite{CBM:2016kpk} and HIAF are being constructed to study heavy-ion collisions with a center-of-mass energy of several GeV for this purpose.
The higher-order moments of net-proton multiplicity and the light nucleus production yield ratios are considered to be sensitive observables to the critical point~\cite{Stephanov_2011,PhysRevD.110.094006}.
With the good detector acceptance and identification ability of charged particles and light nuclei, HHaS can contribute to the global effort of mapping out the QCD phase diagram.

Hypernuclei allow the study of hyperon-nucleon interactions, providing key insights into the strong interaction and strange quark behavior in nuclear environments, which is crucial for the understanding of neutron star interiors where hyperons may exist~\cite{RN425,RN691}.
Heavy-ion collisions at the HIAF energy can create a lot of hyperons, which may coalesce with nucleons in the collision system and form hypernuclei~\cite{RN9}.
The exotic hypernuclei with multistrangeness might be created via HIAF energy heavy-ion collisions with the production cross sections above 1 $\mu$b~\cite{Fe20}. 
With the intense heavy-ion beam at HIAF, various properties of hypernuclei can be measured precisely and new (multi-strange) hypernuclei may be discovered at HHaS.
The hyperon-nucleon interaction in the dense matter can also be studied via the hyperon collective flows and phase-space distribution in heavy-ion collisions at HIAF energies~\cite{Fe24, We24}.

\section{Requirements}
\label{sec:requirements}

To achieve the wide range of physics goals mentioned above, the following key requirements are raised for the design of HHaS.

First, it is required to record the data of collision events with a rate far beyond existing similar experiments.
History has repeatedly demonstrated that new discoveries and understandings may be achieved with increased datum statistics and improved precision.
Taking advantage of the high beam intensity of HIAF, a capability of working with an ultra-high event rate could be a key strong point of HHaS.
HIAF can accelerate a bunch of 2$\times10^{12}$ protons within 0.3 s, which can be slowly extracted to the experimental terminals.
Assuming 1$\%$ of the beam particles react with the target, a collision rate of $\sim$10 GHz can be reached, which is far beyond the event rate of all the currently running experiments.
The bottleneck for acquiring high statistics data is not the beam intensity of HIAF, but the capability of the detector setup to record the events with a high collision rate.
Even the secondary $\pi$ ($K$) beam can reach the intensity of the order of $10^8$ ($10^6$) /s, leading to the collision rate of $\sim 1$ MHz ($\sim 10$ kHz) if the same thickness of a target is assumed.
Unlike very high energy experiments, the triggers using some specific characteristics (e.g., a pair of leptons, a high-energy gamma or jet), which can be easily picked up by a fast detector, cannot cover all the events that contain the broad interesting physics topics of HHaS described in Section~\ref{sec:physics}.
Thus, it is necessary as a good strategy to require all the detectors of HHaS to record signals from all collisions with a high event rate, and then reconstruct, select and analyze the events for interesting physics subjects afterwards.
This is achievable with the state-of-the-art detector technologies, as described in Section~\ref{sec:design} and ~\ref{sec:performance}.

Second, HHaS need to have a large angular, momentum and energy acceptance because it attempts to cover a very wide range of physics studies.
Many physics topics require the reconstruction of a particle with many daughters, or the exclusive measurement of a process with many final particles.
Thus enlarging the acceptance is important for HHaS, in order to carry out the targeted measurements efficiently.

Third, HHaS is required to have good momentum, energy and track pointing resolutions.
A percent level track momentum resolution and several percent level electromagnetic energy resolution should be achieved, in order to obtain a good invariant mass resolution for decayed hadron states.
This is critical to distinguish signals from different light hadron states, which can be closely distributed in mass, and observe a weak exotic signal with a low production rate.
A good track pointing resolution enables HHaS to reconstruct secondary decay vertices precisely, for example, in order to distinguish strange particles and hypernuclei, which usually decay several centimeters away from the collision point, from backgrounds formed by the random combination of final state particles.

Fourth, HHaS should be able to identify many types of final state particles from the collisions, including $e^{\pm}$, $\gamma$, $\pi^{\pm}$, $K^{\pm}$, $p$, $\bar{p}$, deuteron ($d$), triton ($t$), $^3He$ and $^4He$.
The $e^{\pm}$ and $\gamma$ are important final particles for many physics studies.
For example, they are among the daughter particles for almost all the $\eta$ meson decay channels to be studied at HHaS.
Yet, in the hadronic collision environment, their production yield is orders of magnitude lower than hadrons.
Thus very clear identifications of $e^{\pm}$ and $\gamma$ are critical for HHaS to work as an $\eta$ factory experiment.
The light nuclei, i.e., $d$, $t$, $^3He$ and $^4He$, which play an important role in the study of nuclear matter phase transition and hypernuclei, also need to be identified clearly from other particles at HHaS.

Finally, as an advantageous next-generation spectrometer, HHaS is required to have a high single track efficiency.
A minor efficiency loss for each particle could cause a large loss in efficiency for the final measurement which involves numerous particles.
Furthermore, the systematic uncertainties associated with single track efficiencies will also accumulate, leading to a large total systematic uncertainty on the physics measurement.
This can be a critical challenge for HHaS, which targets a new level of the precision for a wide range of physics measurements.
Thus a high track efficiency at 99$\%$ level is essential.

\section{Conceptual Design}
\label{sec:design}

\begin{figure}[!htb]
\includegraphics
  [width=0.98\hsize]  {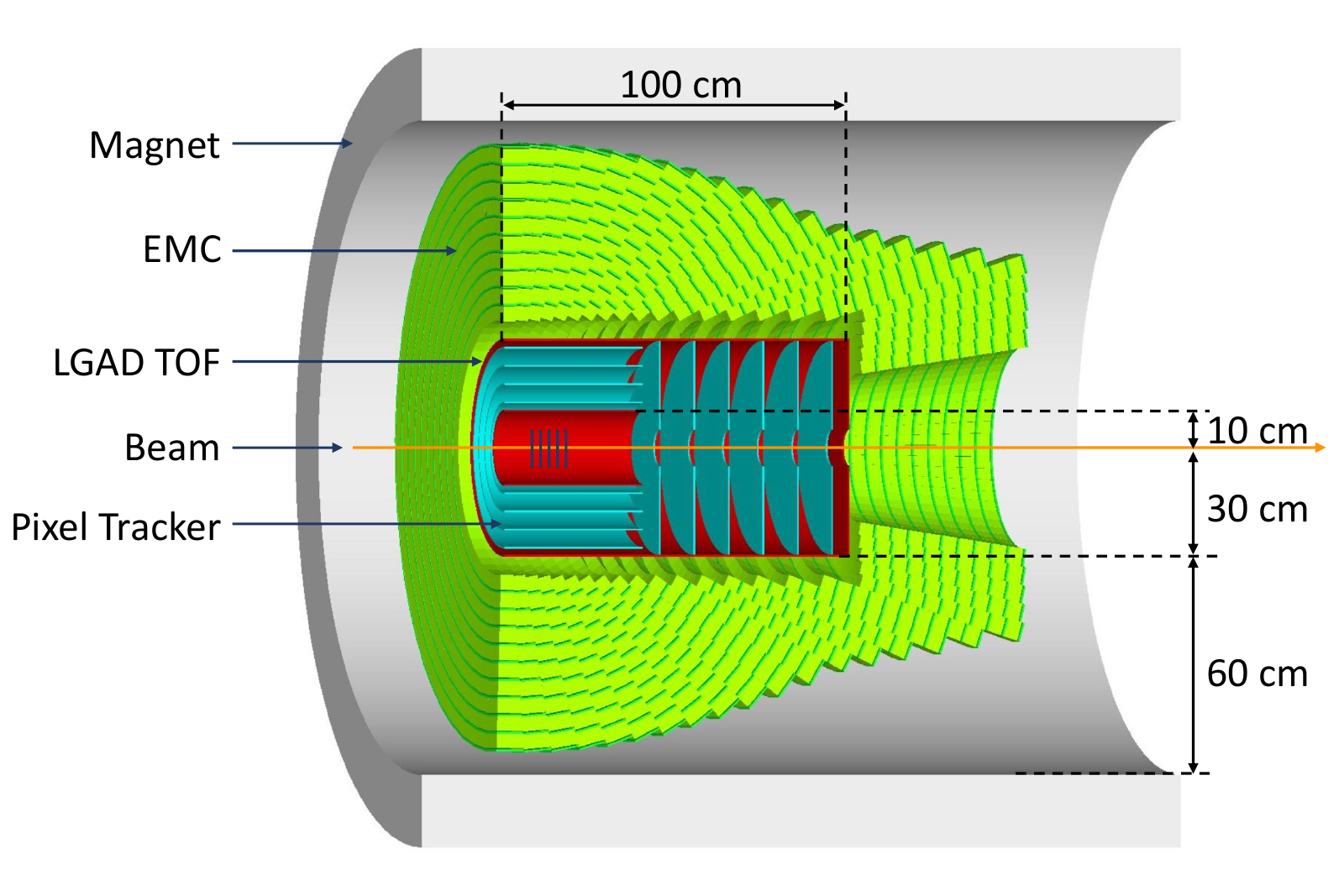}
\caption{HHaS conceptual design.}
\label{fig:HHaS}
\end{figure}

The detector system of HHaS comprises a solenoidal magnet, a pixel tracker, a time-of-flight (TOF) detector and an electromagnetic calorimeter (EMC) that work in concert to meet the above-mentioned requirements, as shown in Fig.~\ref{fig:HHaS}. 

A solenoidal magnet that provides a 1.5-4 Tesla field forms the backbone of the spectrometer, enabling the precise momentum measurement of charged particles. Depending on the physics requirements, the magnetic field can be lowered to detect particles with low transverse momenta ($p_{\rm T}$), or increased to achieve better momentum resolutions. 

The core tracking detector utilizes the advanced Monolithic Active Pixel Sensors (MAPS) technology~\cite{SNOEYS2014167,CONTIN20161155,Kugathasan:2019ydo,AGLIERIRINELLA2024169896}, which provides excellent position resolution, fast readout, and moderate material budget~\cite{TURCHETTA2001677,TURCHETTA2006139,CONTIN201860}. 
The pixel sensors are arranged in cylindrical barrel layers surrounding the target, together with disks covering the forward regions, enabling the detection of particles across a wide range of the polar angle from 10 to 100 degrees with the full azimuthal coverage.
Unlike traditional MAPS sensors, which detect only whether a pixel is hit, the MAPS sensors developed for HHaS are designed to simultaneously measure both time information and the magnitude of ionization charge for each pixel.
This design allows the HHaS pixel tracker to function as an innovative five-dimensional tracking device, capable of synchronously acquiring three-dimensional spatial coordinates, hit timings, and ionization charge amounts.
The time measurement with a resolution of approximately 10 ns enables distinguishing hits from different collisions when HHaS operates at an ultra-high event rate.
The ionization charge measurement helps distinguish different nuclei with identical charge-to-mass ratios (e.g., d, $^4He$, and $^6Li$) based on the differences in their energy deposition. 
This capability is crucial for studying hypernuclei and nuclear matter at high baryon densities.
To accommodate the increased power consumption and technical challenges posed by time and ionization charge measurements, the MAPS sensors for HHaS have a moderate pixel pitch of approximately 100 $\mu$m, larger than the pixel pitches of MAPS sensors used in higher energy experiments focusing on heavy-flavor particle's decay measurements~\cite{2012A, MAGER2016434}. 
By using the ionization charge as a weight to calculate the center of gravity of each hit cluster, a hit position resolution of 30 $\mu$m or better can be achieved, which meets the requirements of HHaS, as demonstrated in Section~\ref{sec:performance}. 
With the excellent position resolution of the pixel tracker and the relatively strong magnet field, HHaS can reconstruct and measure charged particles with minimal track lengths.
Thus, HHaS is designed as a very compact and cost-effective spectrometer, as presented in Fig.~\ref{fig:HHaS}.
Considering a typical sensor detection efficiency of 99$\%$ and dead area of ~1$\%$ between neighboring sensors, the hit efficiency per layer should be about 98$\%$.
A track efficiency over 99$\%$ can be achieved by allowing at most one missed hit across all layers that a track penetrates.
Thanks to the high granularity of the pixel detector, the hit and track efficiency losses due to the detector unit occupancy caused by a high particle multiplicity or a high event rate are limited, as discussed in Section ~\ref{sec:performance}.
With carbon fiber supporting structure, flexible PCB cables and air cooling, the material budget can be constrained down to $X/X_{0}\sim0.3\%$ per layer, so that the multiple coulomb scattering effect of particles is at an acceptable level.
Currently, significant R\&D efforts are focused on developing the Nupix-H~\cite{10689670}, a series of MAPS sensors tailored for HHaS applications. 

Inside and outside of the pixel tracking detector, the Capacitively-coupled (AC) Low-Gain Avalanche Detectors (LGAD) ~\cite{TORNAGO2021165319, SUN2024169203} are placed for TOF measurements.
The average radii for the inner and outer barrel layers are 10 and 30 cm, respectively.
A disk layer is placed to cover the forward direction.
The outer layers measure the flight end time of each charged particle, 
while the collision time, i.e., the common start time of all primary particles, is measured by the inner LGAD layer using all charged particles that are detected there.
A remarkable time resolution of about 30 ps can be achieved with the state-of-the-art LGAD technology, which helps both the identification of various charged particles and the separation of tracks from different collisions.
A charge-weighted hit position resolution below 30 $\mu$m can be obtained with a r$\times\phi$ strip pitch around 100 $\mu$m, so that the LGAD TOF provides another 1 or 2 precise hits in addition to the pixel tracker and a longer track length for momentum reconstruction.
The z strip pitch is much larger at cm scale in order to have acceptable channel density, total layer material thickness and power consumption.
With this design, a material budget of $X/X_{0}\sim3\%$ per layer is expected.
Unlike Direct-coupled (DC) LGAD, the AC LGAD has no dead area surrounding each pixel or strip, thus can achieve a high hit efficiency above 98$\%$.
The efficiency loss induced by the detector unit occupancy is also very small as described in Section ~\ref{sec:performance}.

The dual-readout EMC will be installed outside the LGAD TOF detector.
As shown in Fig.~\ref{fig:HHaS}, layers of lead glass and plastic scintillator are assembled in an alternating sequence, forming towers pointing to the target area, with a total thickness of X/X$_0$ $\sim$ 14.
Silicon photomultipliers (SiPMs) are coupled directly to the lead-glass and plastic-scintillator tiles to read out Cherenkov and scintillation light, respectively, thereby enabling effective separation between electromagnetic and hadronic showers.
This separation is possible because (1) most particles in a hadronic shower with  energies around 1 GeV lack sufficient velocity to produce Cherenkov radiation
and (2) electrons in an electromagnetic shower can easily produce Cherenkov light due to their sub-MeV/$c^{2}$ mass ~\cite{Gatto:2022bzg,Akchurin:2014zna}.
This type of dual-readout EMC, which is named ADRIANO-II, has also been adopted by the REDTOP collaboration~\cite{REDTOP:2022slw,2799170}.
Tests by the REDTOP collaboration show that ADRIANO-II can achieve an energy resolution of approximately 3\% at 1 GeV. 
Our simulations indicate the performance comparable to these tests (see Section~\ref{sec:performance}).
A time resolution of approximately $200~\text{ps}$ facilitates the separation of signals from different events.
The module dead time must be at the microsecond level so that the data acquisition efficiency should be reasonably high when HHaS operates at high event rates.

Depending on the physics topics to be studied, a multi-layer solid nuclear target, a liquid hydrogen target, or a polarized $^3He$ target may be placed in the center of HHaS along the beamline.
The polarized $^3He$ target can resemble a polarized neutron target after subtracting the contribution from the beam colliding with the protons in it.
With the innermost LGAD layer 10 cm from the beamline, enough space is left to accommodate the usage of all these types of targets, especially the liquid hydrogen and polarized $^3He$ targets, which have a dewar to work at a low temperature.

\section{Expected Performance}
\label{sec:performance}
The capability to handle an ultra-high event rate is an advantage that sets HHaS apart from many existing experimental setups.
For low multiplicity collisions with a proton beam, HHaS can take 100 M events per second.
Due to the choice of fast detector technologies for all the three detectors, HHaS runs without triggers and reads out all signals that the detectors receive. 
Then, using the read-out time information, signals can be sorted out as belonging to different events.
With an event rate of 100 MHz, there are only 10 collision events on average in a $\pm$5$\sigma$ ($\sim$100 ns) time window of the pixel tracker.
The total number of charged tracks is less than 100 from these $\sim$10 events when using a proton beam, which can be well handled by the track finding algorithm.
After that, tracks from the $\sim$10 different collisions are further separated using the time information of the 1 or 2 LGAD hits in each track.
As the 30-ps time resolution of LGAD is much more precise compared to the average collision time interval of 10 ns, tracks from different collisions can be separated unambiguously.
The EMC time resolution of about 200 ps also allows distinguishing clearly signals from different collisions.

Another challenge coming along with the ultra-high event rate is the occupancy of detector units.
The pixels, LGAD strips and EMC towers are read out independently.
For example, when one pixel is hit, it takes about 10 $\mu$s to read out the time and charge of this hit.
This pixel is dead during this time, but other pixels remain ready to detect signals from other particles.
The detector occupancy, i.e., portion of dead units during data taking, is proportional to the unit dead time, particle multiplicity of a collision event and event rate, and is inverse-proportional to the granularity of the detector.
Considering that a typical p + $^7Li$ collision at 1.8 GeV generates 4 charged particles and 0.4 $\gamma$, with the pixel, LGAD strip, and EMC tower read-out dead time of 10, 2 and 1 $\mu$s, respectively, the average occupancies of the innermost pixel layer, the outer LGAD layer and the EMC are estimated to be 0.02$\%$, 0.1$\%$ and 7$\%$, respectively, with an event rate of 100 MHz.
Thus, the efficiency loss due to detector occupancy is acceptable.
For collisions with a heavy ion beams, hundreds of particles per event can be generated, thus HHaS runs at a lower event rate around 1 MHz in order to avoid a high detector occupancy and a large efficiency loss.

The radiation dose and 1 MeV neutron equivalent fluence for various materials and components of HHaS have been simulated with the FLUKA software~\cite{Ferrari:2005zk}. 
The result shows that most detector components can sustain the radiation for 1 year of accumulated running time. 
However, the innermost lead glass tiles of the EMC will receive a radiation dose up to 500 Gy. There are only a limited number of references about radiation hardness for the broad range of existing lead glass types. 
Thus, detailed measurements need to be carried out to select a lead glass type that can sustain such a high radiation dose.

With the 1-100 MHz event rate, HHaS is several orders of magnitude faster than most of the existing high energy particle and nuclear physics experiments, granting HHaS a great potential for discoveries and precision measurements.
The event rates with secondary $\pi$ and $K$ beams are at 1 MHz and 10 kHz levels, respectively, which are limited by the beam intensities as mentioned in Section~\ref{sec:requirements}.

HHaS accepts charged particles with the $p_{\rm T}$ above 50 MeV/$c$ (in case of the magnetic field of 1.5 T) and $\gamma$ with the energy above 50 MeV across a wide angular range from 10 to 100 degrees in the polar angle ($\theta$) with the full azimuthal coverage.
Some low-$p_{\rm T}$ tracks may not be able to reach the outer barrel LGAD layer, but can swirl in-and-out several loops until hitting the endcap LGAD disk.
A successful momentum reconstruction and TOF measurement can still be achieved for these low-$p_{\rm T}$ tracks, at least in low multiplicity events using $p$, $\pi$ and $K$ beams.
This $p_{\rm T}$, energy and angle acceptance is very large for particles produced by a multi-GeV beam bombarding a fixed target, making HHaS a very efficient experiment for exclusive reaction channel measurements and for the reconstruction of particles that decay into multiple daughters.

\begin{figure}[!htb]
\includegraphics
  [width=0.98\hsize]
  {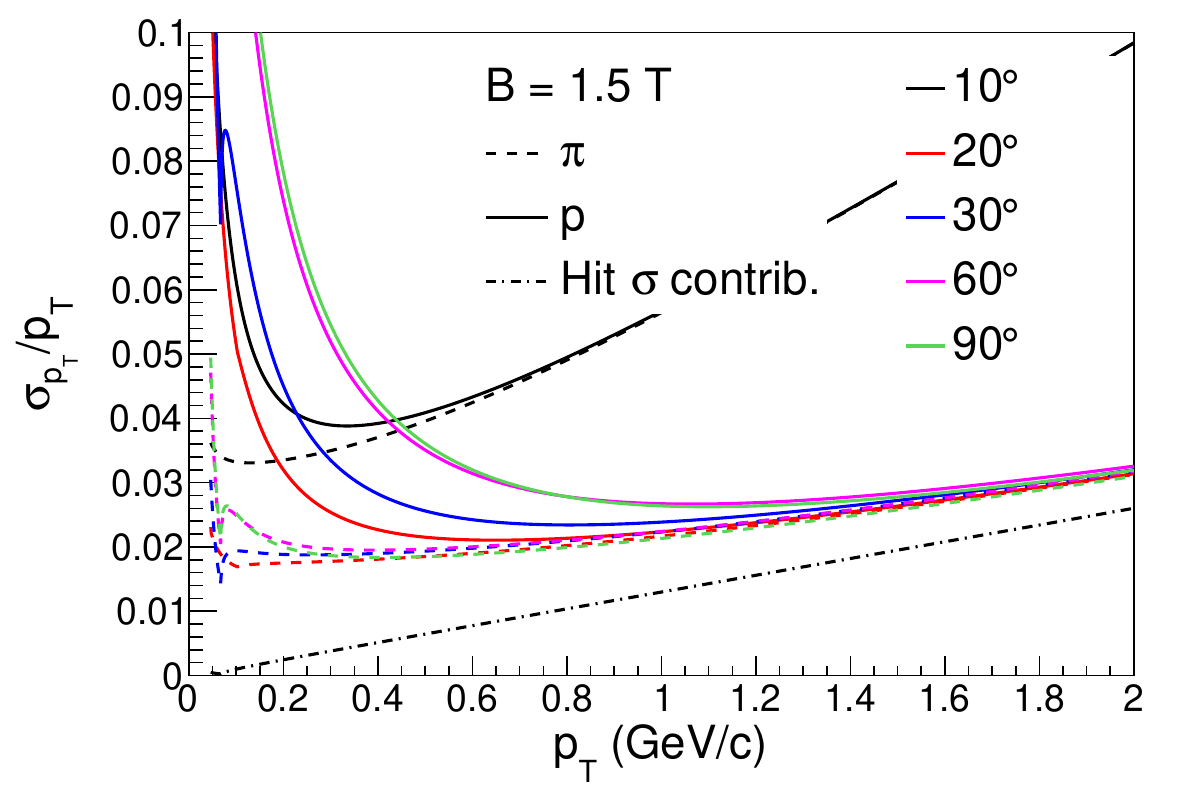}
\includegraphics
  [width=0.98\hsize]
  {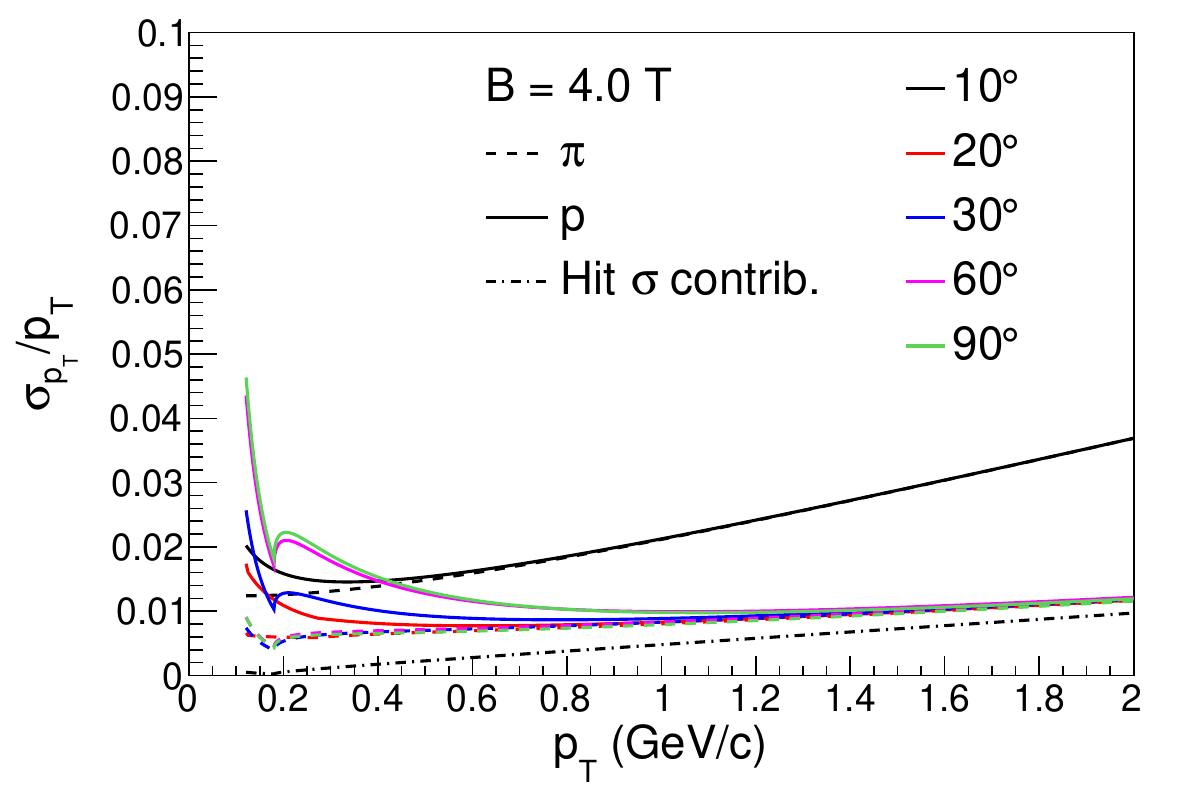}  
\caption{Estimated momentum resolution as a function of transverse momentum for $\pi^{\pm}$ and protons at different polar angles, with 1.5 T (upper panel) and 4 T (lower panel) magnetic fields. The $p_{\rm T}$ resolution from hit position resolution alone is shown by the dashdotted line.}
\label{fig:pResolution}
\end{figure}

The hit position resolution of the pixel tracker and the LGAD TOF (in the r$\times\phi$ direction) is around or below 30 $\mu$m. 
Thanks to the good hit precisions, HHaS can reconstruct charged particle tracks with good momentum resolutions, even though the radius of the outer LGAD TOF layer is only 30 cm.
Fig.~\ref{fig:pResolution} shows the estimated momentum resolution as a function of $p_{\rm T}$ for $\pi^{\pm}$ and protons at different polar angles, assuming a 30-$\mu$m hit-position resolution and a pixel detector thickness of X/X$^0$ = 0.3$\%$ per layer.

For particles with the $p_{\rm T}$ below 1 GeV/c, the dominant contribution of the $p_{\rm T}$ resolution is the multiple Coulomb scattering effect, which decreases with increasing $p_{\rm T}$.
The hit resolution contribution, shown by the dashdotted line, increases linearly with $p_{\rm T}$ and becomes important only at a very large $p_{\rm T}$, where the particle production yield is very low.
There is a kink structure with $p_{\rm T}$ = 68 (180) MeV for the B = 1.5 (4) T case, because that's the point where the tracks are barely able to reach the outer barrel LGAD layer.
Below this $p_{\rm T}$, as the transverse track length increases with increasing $p_{\rm T}$, the $p_{\rm T}$ resolution also improves rapidly. 
Above this $p_{\rm T}$, the track becomes straighter and shorter as $p_{\rm T}$ increases, and the $p_{\rm T}$ resolution gets worse a bit with increasing $p_{\rm T}$, until the decrease of the multiple Coulomb scattering effect becomes dominant.

For most charged particles, a momentum resolution of around 1$\%$ can be achieved with a magnetic field of 4 T, except for very forward particles (e.g., $\theta$ = 10$^\circ$) with a limited transverse track length or very low momentum particles which suffer from a large multiple Coulomb scattering effect.
The use of a lower magnetic field strength of 1.5 T degrades the momentum resolution to a level of 3$\%$ but improves the acceptance for low $p_{\rm T}$ particles.

\begin{figure}[!htb]
\includegraphics
  [width=0.9\hsize]
  {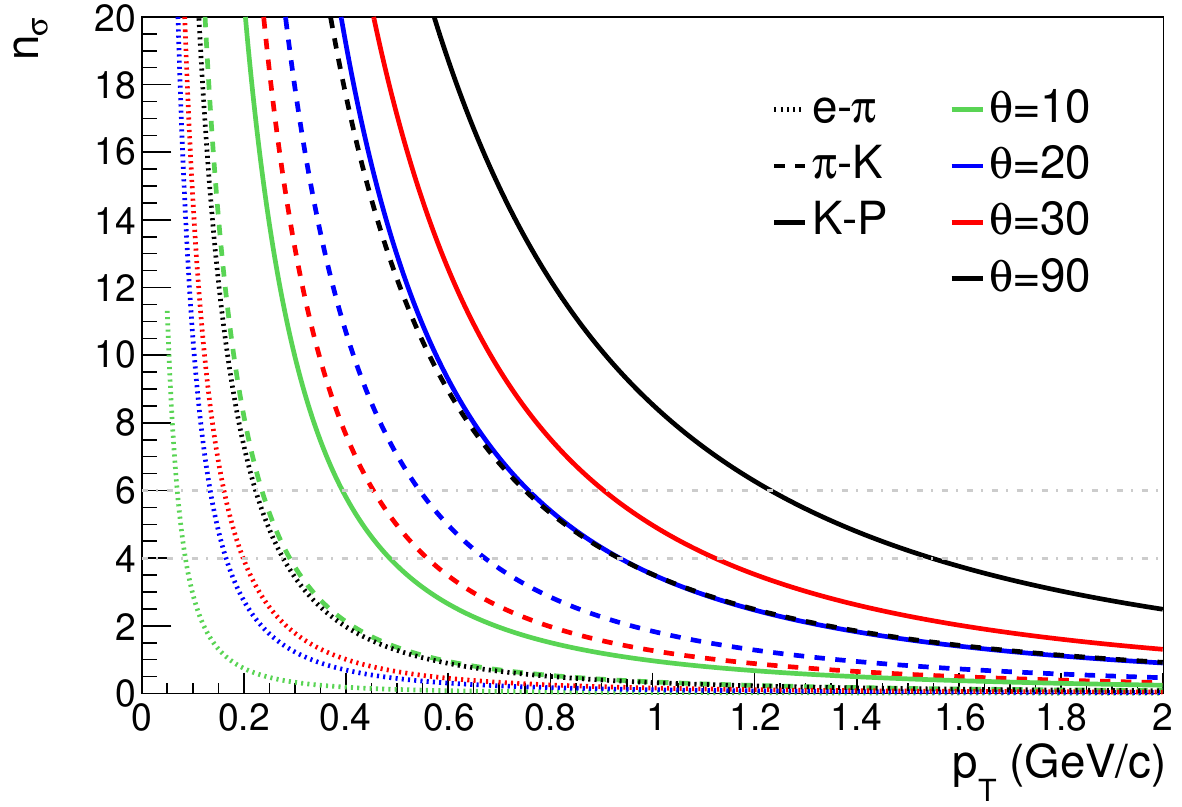}
\caption{n$_\sigma$ separation as a function of transverse momentum for various charged particles at different polar angles using TOF measurements.}
\label{fig:TofPidNSigma}
\end{figure}

Once the momenta are measured, charged particles are mainly identified by the measured TOF.
Assuming LGAD TOF time resolution of $\delta$t = 30 ps, the separation ability for various charged particles at different polar angles ($\theta$) is shown in Fig.~\ref{fig:TofPidNSigma}. 

\begin{figure}[!htb]
\includegraphics
  [width=0.98\hsize]
  {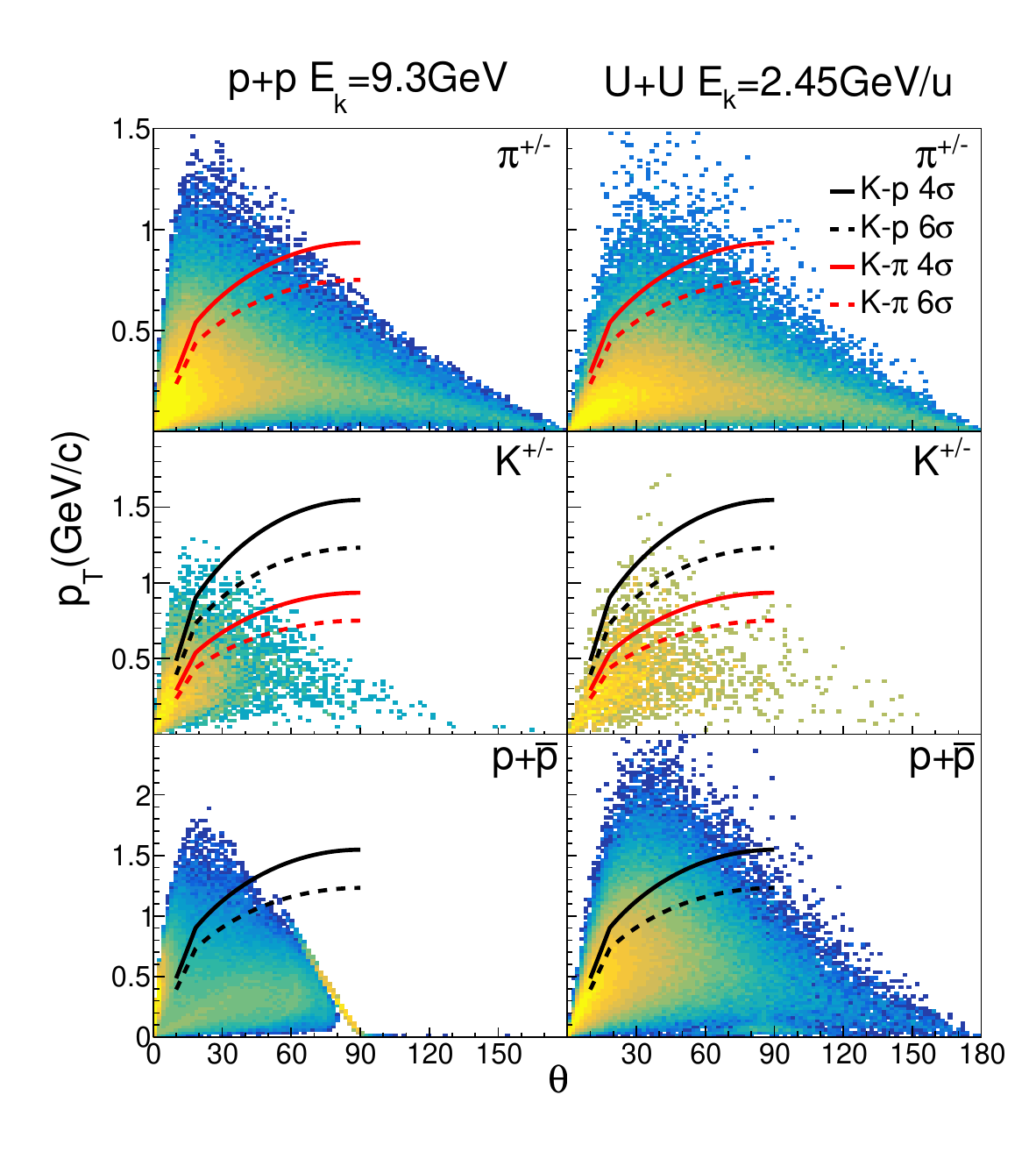}
\caption{$p_{\rm T}$ vs. $\theta$ distributions for $\pi^{\pm}$, $K^{\pm}$ and proton produced with top-energy proton and uranium beams at HIAF from GiBUU~\cite{Buss:2011mx} simulation, comparing with the $p_{\rm T}$ for 4-$\sigma$ and 6-$\sigma$ $\pi$-K and K-p separation.}
\label{fig:TofPidAndParticleDistribution}
\end{figure}

In Fig.~\ref{fig:TofPidAndParticleDistribution}, the $p_{\rm T}$ for 4-$\sigma$ and 6-$\sigma$ $\pi$-K and K-p separation is plotted as a function of $\theta$ on top of the produced $\pi^{\pm}$, $K^{\pm}$ and (anti-)proton distributions for the cases of top-energy proton (9.3 GeV) and uranium (2.45 GeV/u) beams at HIAF.
It can be seen that the majority of the produced charged particles sit below the $\pi$-K and K-p separation lines, meaning that they can be well separated with the TOF measurements.
It should also be noted that the top energy collisions are the most difficult cases for charged particle identification.
With lower beam energies, the produced particles will be distributed at larger $\theta$, and the mean $p_{\rm T}$ will decrease slightly, making the particle identification easier.

\begin{figure}[!htb]
\includegraphics
  [width=0.8\hsize]
  {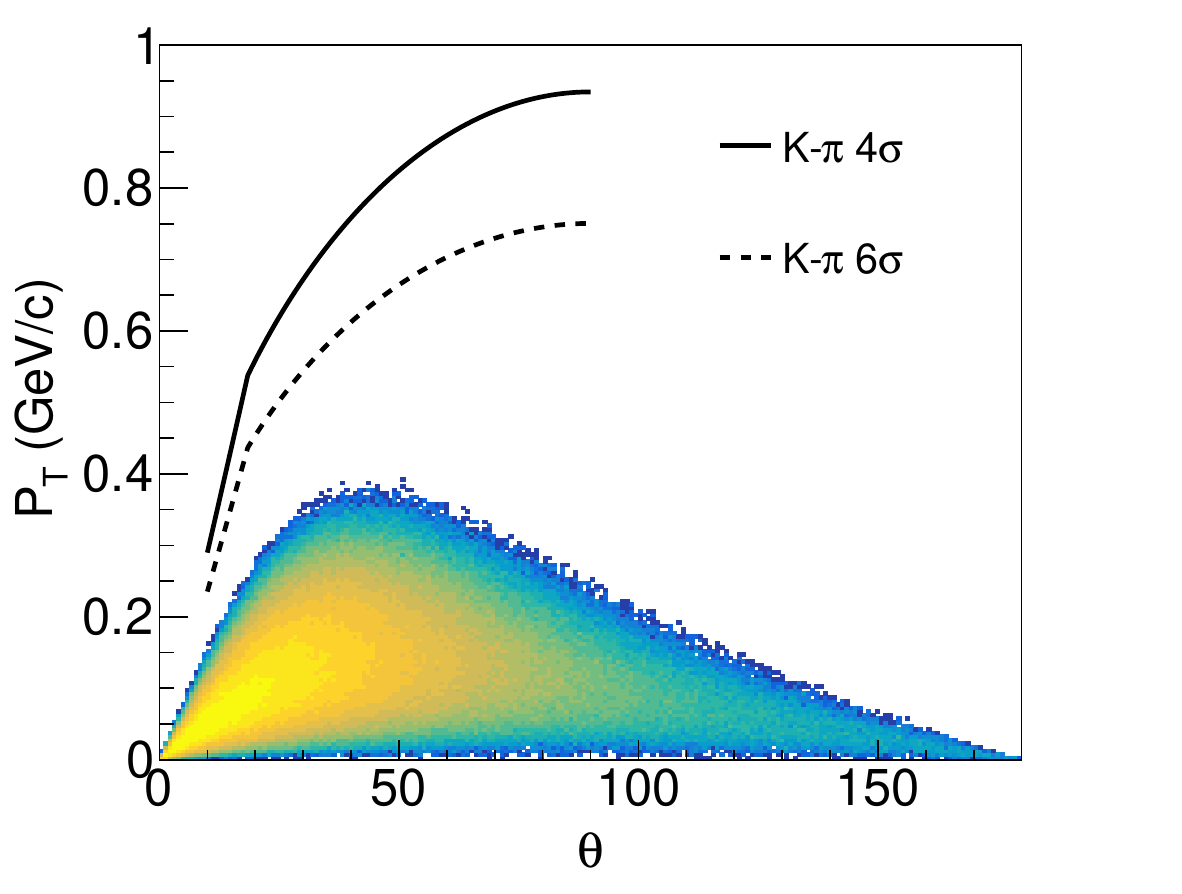}
\caption{$p_{\rm T}$ vs. $\theta$ distribution for $\pi^{\pm}$ from $\eta\rightarrow\pi^{+}\pi^{-}\pi^{0}$ decay, for the $\eta$ meson produced by a proton beam with a kinetic energy of 1.8 GeV bombarding a $^{7}Li$ target, comparing with the $p_{\rm T}$ for 4-$\sigma$ and 6-$\sigma$ $\pi$-K separation.}
\label{fig:etaDecayedPiDistribution}
\end{figure}

For the search and measurement of decayed particles, the daughter particles tend to have lower $p_{\rm T}$'s than primary particles, which dominates the distributions in Fig.~\ref{fig:TofPidAndParticleDistribution}.
Fig.~\ref{fig:etaDecayedPiDistribution} shows the $p_{\rm T}$ vs. $\theta$ for the $\pi^\pm$ mesons from the $\eta\rightarrow\pi^{+}\pi^{-}\pi^{0}$ decay in the case that the $\eta$ meson is produced by a proton beam with a kinetic energy of 1.8 GeV bombarding a $^{7}Li$ target.
As shown in the plot, the 4-$\sigma$ and 6-$\sigma$ $\pi$-K separation curves are well above the $\pi^{\pm}$ distribution, meaning that all the $\pi^{\pm}$ from the $\eta$ decay can be well identified.

\begin{figure}[!htb]
\includegraphics
  [width=0.98\hsize]
  {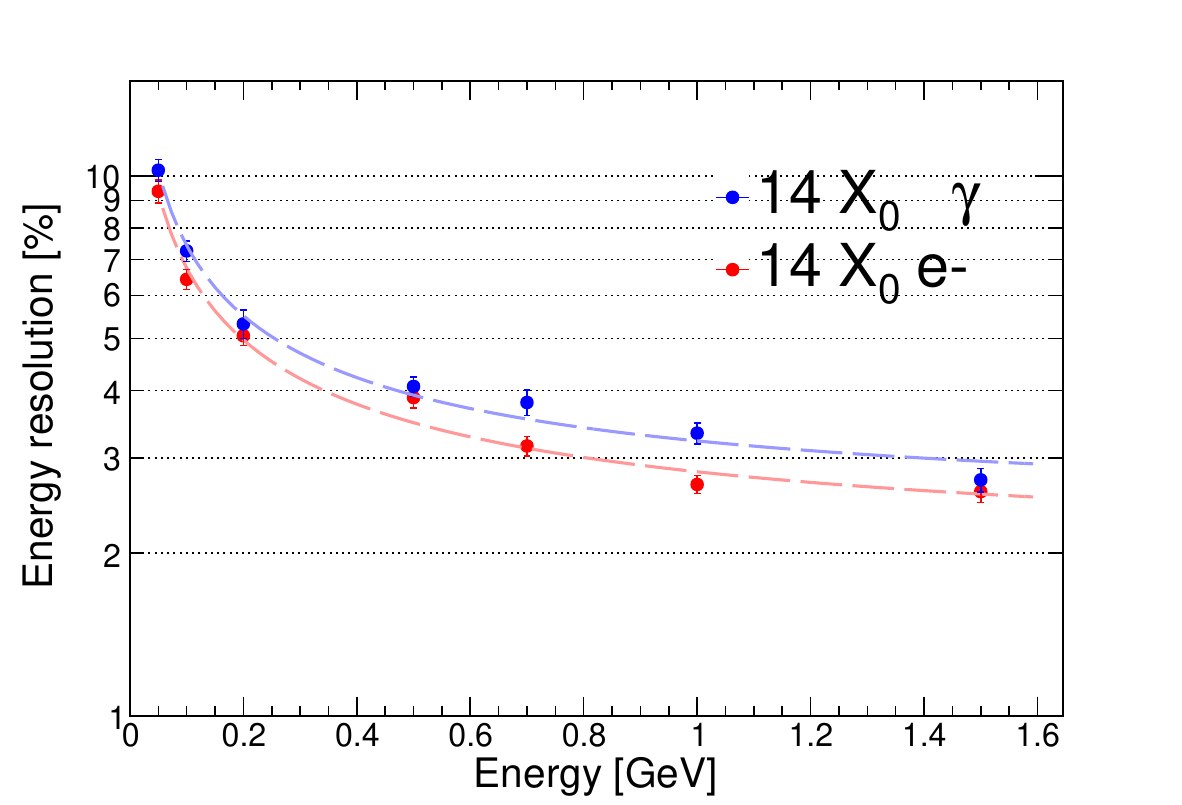}
\caption{Energy resolution as a function of energy for electrons and $\gamma$ from MC simulations.}
\label{fig:EResolution}
\end{figure}

The response of the dual-readout EMC to electrons and $\gamma$ are simulated using the Geant4 package~\cite{geant1, geant2, geant3}. 
The result of electromagnetic energy ($E_{EM}$) resolution is shown in Fig.~\ref{fig:EResolution}. 
An energy resolution of $\sim$3$\%$ @ 1GeV can be obtained.
For the ADRIANO-II type calorimeter, most of the energy loss happens in the denser and thicker lead glass, where only Cherenkov light is generated.
The number of Cherenkov photons generated per unit energy loss is orders of magnitude lower than that of scintillation photons.
Thus, the total number of generated Cherenkov plus scintillation photons in the ADRIANO-II EMC is much less than that in a traditional EMC by inorganic scintillators~\cite{INFNRD-FA:2020fiu}.
However, due to small tile sizes, the photons can hit the SiPM's before traveling a very large distance.
This helps the ADRIANO-II EMC to gain a better photon collection efficiency and obtain an energy resolution close to the expensive scintillating crystal calorimeters.

\begin{figure}[!htb]
\includegraphics
  [width=0.98\hsize]
  {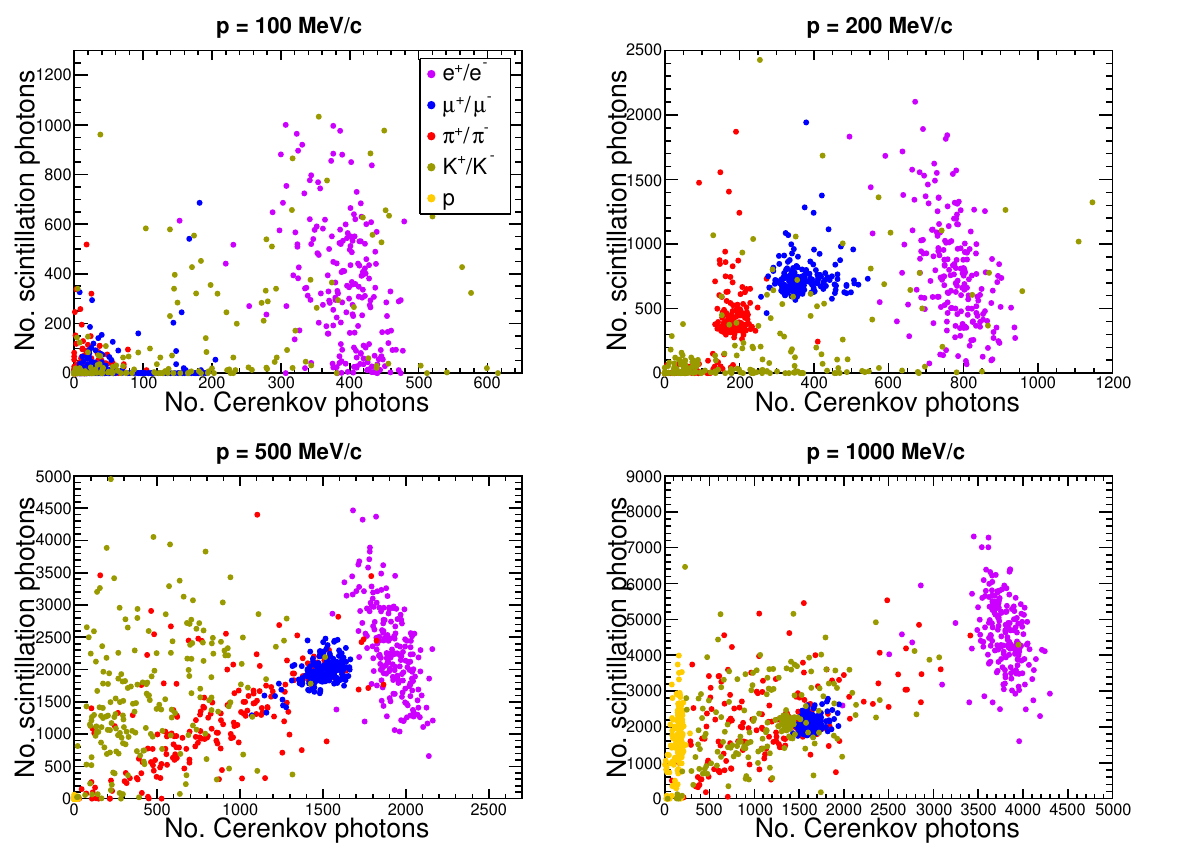}
\caption{Numbers of scintillation vs. Cherenkov photons detected by the SiPM's in the dual-readout EMC for various charged particles with certain momenta from Geant4 simulations.}
\label{fig:EMC_PID_charged}
\end{figure}

\begin{figure}[!htb]
\includegraphics
  [width=0.8\hsize]
  {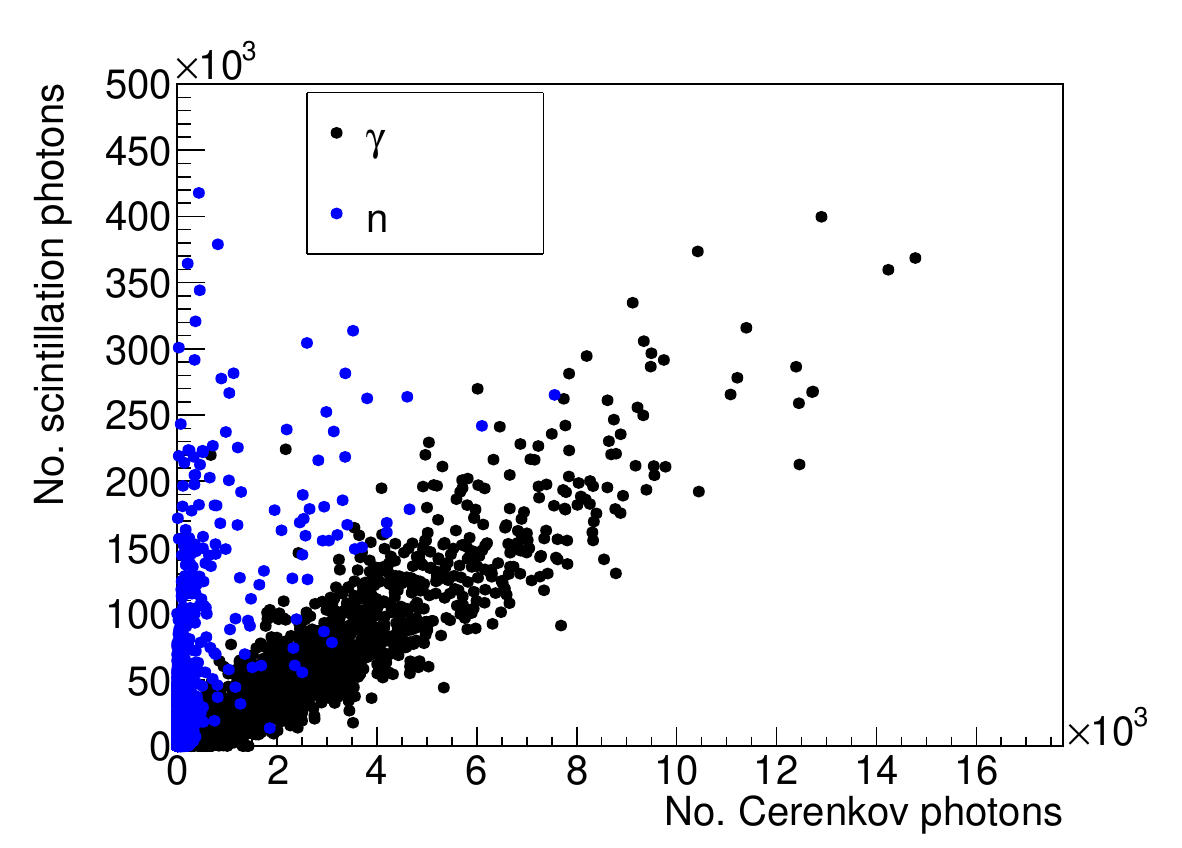}
\caption{Numbers of scintillation vs. Cherenkov photons detected by the SiPM's in the dual-readout EMC for $\gamma$ and neutrons from Geant4 simulations.}
\label{fig:EMC_PID_uncharged}
\end{figure}

The Cherenkov-scintillation dual-readout EMC has a very good ability to separate electromagnetic shower signals from the backgrounds generated by hadrons.
Figure~\ref{fig:EMC_PID_charged} illustrates how electrons possessing certain momenta can be identified clearly from other charged particles using the number of scintillation photons vs. Cherenkov photons measured by the EMC.
In contrast, the separation of uncharged $\gamma$'s and neutrons is more challenging, because without the particle momentum measurement from the tracker, a $\gamma$ with a certain energy can be contaminated by a higher energy neutron, which leaves a part of its energy in the calorimeter~\cite{Pezzotti:2022ndj, Gatto:2024dmg, Akchurin:2009zzb}.
Figure~\ref{fig:EMC_PID_uncharged} shows the numbers of scintillation vs. Cherenkov photons detected by the SiPM's in the dual-readout EMC for $\gamma$'s and neutrons from MC simulations.
The $\gamma$ sample used here comes from the $\eta\rightarrow\pi^+\pi^-\pi^0\rightarrow\pi^+\pi^-\gamma\gamma$ decay with a 1.8 GeV proton beam bombarding a $^7Li$ target, simulated with the GiBUU model~\cite{Buss:2011mx}.
The background neutron sample is generated with the same collision system from GiBUU simulation.
Good $\gamma$-n separation is achieved except for very low-energy $\gamma$'s distributed at the bottom left corner of the plot.
Considering the relatively good energy resolution, the excellent electromagnetic-hadronic shower separation ability and the very low cost of lead glass and plastic scintillator, ADRIANO-II is a very suitable and cost-effective EMC for HHaS.

\begin{figure}[!htb]
\includegraphics
  [width=0.98\hsize]
  {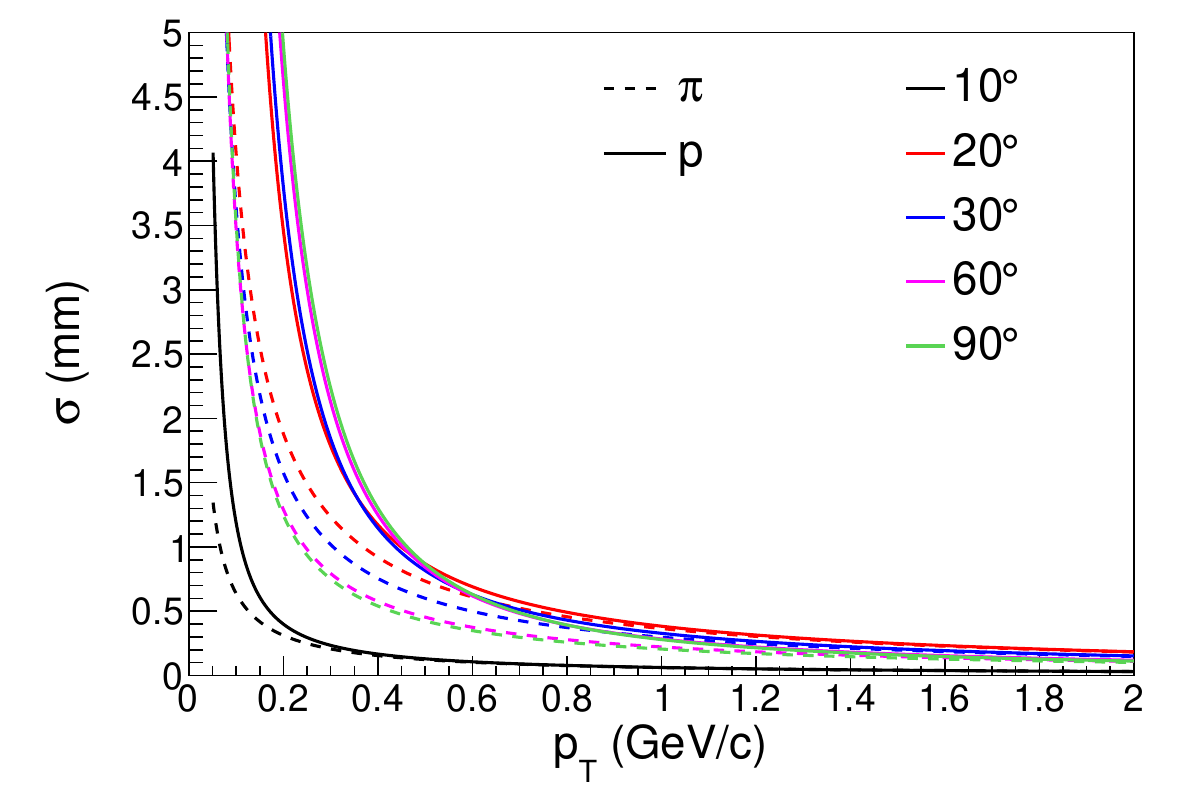}
\caption{Estimated track pointing resolution around the collision point as a function of $p_{\rm T}$ for $\pi^{\pm}$ and protons at different polar angles.}
\label{fig:pointingResolution}
\end{figure}

With the advantage of the good hit resolution of the pixel tracker, HHaS is expected to have a good track pointing resolution around the collision point.
The dominant contribution to the track pointing resolution is the multiple Coulomb scattering effect.
Figure~\ref{fig:pointingResolution} shows the estimated track pointing resolution as a function of $p_{\rm T}$ for $\pi^{\pm}$ and protons at different polar angles.
The pointing resolution for tracks with a polar angle of 10 degrees is better than others because they do not penetrate the inner LGAD barrel, which is much thicker than the pixel layers. 
For protons with a typical $p_{\rm T}$ of 500 MeV/c, the pointing resolution is around 0.9 mm.
For $\pi^{\pm}$ with a $p_{\rm T}$ of 200 MeV/c, the pointing resolution is about 1.5 mm.
These mm level pointing resolutions allow HHaS to reconstruct secondary decay vertices precisely, so that particles with strangeness and hypernuclei, which have a typical decay c$\tau$ of several cm, can be clearly identified from combinatorial backgrounds.

Table~\ref{tab:performance} summarizes the expected performance of HHaS, which are described above.

\begin{table}[!htb]
\caption{Summary of the performance expected for HHaS.}
\label{tab:performance}
\begin{tabular*}{0.96\hsize} {@{\extracolsep{\fill} } lr}
\toprule
Event rate & $\sim$100 MHz (p beam) \\
& $\sim$1 MHz (heavy-ion beam) \\
& $\sim$1 MHz ($\pi$ beam) \\
& $\sim$10 kHz ($K$ beam) \\
Angular acceptance & 10$^{\circ}$ < $\theta$ < 100$^{\circ}$ \\
& 0 < $\phi$ < 2$\pi$ \\
$p_{\rm T}$ range & > 50 MeV/$c$ (B = 1.5 T) \\
& > 130 MeV/$c$ (B = 4 T) \\
$E_{EM}$ range & > 50 MeV \\
Track efficiency & $\sim$99$\%$ \\
TOF efficiency & $\sim$98$\%$ \\
Typical $p_{\rm T}$ resolution & $\sim$3\% (B = 1.5 T) \\
& $\sim$1\% (B = 4 T) \\
$E_{EM}$ resolution & $\sim$3\% @ 1 GeV \\
Track pointing resolution & $\sim$0.9 mm (p @ 500 MeV/c) \\
& $\sim$1.5 mm ($\pi^{\pm}$ @ 200 MeV/c) \\
Identified particles & $e^{\pm}$, $\gamma$, $\pi^{\pm}$, $K^{\pm}$, $p$, $\bar{p}$, $d$, $t$, $^3He$, $^4He$ \\
\bottomrule
\end{tabular*}
\end{table}

Preliminary simulations for the $\eta$ meson physics to be conducted at HHaS are given in Ref.~\cite{Chen:2024wad}. 
With a running time of only one month, HHaS is capable of taking a sample of 5.9$\times 10^{11}$ $\eta$ mesons, which is three orders of magnitude larger than the total $\eta$ samples recorded globally to date.
One to two orders of magnitude higher sensitivities than the current results can be achieved in the search for dark photon and dark Higgs, and the tests of C and CP symmetries. 
This demonstrates the great advantage from the outstanding event rate and the comprehensive capabilities of HHaS.
Detailed simulation studies will be carried out for more potential physics measurements at HHaS.

\section{Summary}

In summary, we propose a new experimental setup, named HHaS, at the HIAF high energy terminal. 
This setup consists of a solenoidal magnet, a 5-dimensional pixel tracking detector employing MAPS technology, a LGAD TOF detector and a Cherenkov-scintillation dual-readout EMC.
HHaS has a large particle acceptance, a typical momentum resolution of $\sim$1\% for charged particles, an EM energy resolution of $\sim$3\% @ 1 GeV and a track pointing resolution at mm level. 
It can clearly identify and measure $e^{\pm}$, $\gamma$, $\pi^{\pm}$, $K^{\pm}$, $p$, $\bar{p}$, $d$, $t$, $^3He$ and $^4He$.
Most impressively, the system can sustain interaction 
rates of $\sim$100 MHz and $\sim$1 MHz for proton and heavy-ion beam experiments, respectively.
This capability sets HHaS apart from many existing experimental setups, making it a very versatile and competitive spectrometer to search for BSM physics, to test fundamental symmetries, to study light hadron spectroscopy and related dynamics, to measure the properties of hypernuclei and probe hyperon-nucleon interactions, to discover exotic hadron states and multi-strange hypernuclei, as well as to map the QCD phase diagram.

\bibliography{references}

\end{document}